\documentstyle[12pt]{article}
\newcommand{\nc}{\newcommand}

\nc{\dbar}{\bar{\partial}}
\nc{\be}{\begin{equation}}
\nc{\ee}{\end{equation}}

\catcode`\@=11

\@addtoreset{equation}{section}
\def\theequation{\thesection\arabic{equation}}

\def\@normalsize{\@setsize\normalsize{15pt}\xiipt\@xiipt
\abovedisplayskip 14pt plus3pt minus3pt%
\belowdisplayskip \abovedisplayskip
\abovedisplayshortskip  \z@ plus3pt%
\belowdisplayshortskip  7pt plus3.5pt minus0pt}
\def\small{\@setsize\small{13.6pt}\xipt\@xipt
\abovedisplayskip 13pt plus3pt minus3pt%
\belowdisplayskip \abovedisplayskip
\abovedisplayshortskip  \z@ plus3pt%
\belowdisplayshortskip  7pt plus3.5pt minus0pt
\def\@listi{\parsep 4.5pt plus 2pt minus 1pt
            \itemsep \parsep
            \topsep 9pt plus 3pt minus 3pt}}

\def\underline#1{\relax\ifmmode\@@underline#1\else
        $\@@underline{\hbox{#1}}$\relax\fi}
\@twosidetrue
\relax

\catcode`@=12

\catcode`\@=11

\def\section{\@startsection{section}{1}{\z@}{3.5ex plus 1ex minus
   .2ex}{2.3ex plus .2ex}{\large\bf}}

\def\ps@headings{\def\@oddfoot{}\def\@evenfoot{}
\def\@oddhead{\hbox{}\hfill
        \makebox[.5\textwidth]{\raggedright\ignorespaces --\thepage{}--
        \hfill }}
\def\@evenhead{\@oddhead}
\def\subsectionmark##1{\markboth{##1}{}}
}

\ps@headings

\catcode`\@=12

\relax

\def\figcap{\section*{Figure Captions\markboth
        {FIGURECAPTIONS}{FIGURECAPTIONS}}\list
        {Fig. \arabic{enumi}:\hfill}{\settowidth\labelwidth{Fig. 999:}
        \leftmargin\labelwidth
        \advance\leftmargin\labelsep\usecounter{enumi}}}
 \relax
\def\tablecap{\section*{Table Captions\markboth
        {TABLECAPTIONS}{TABLECAPTIONS}}\list
        {Table \arabic{enumi}:\hfill}{\settowidth\labelwidth{Table 999:}
        \leftmargin\labelwidth
        \advance\leftmargin\labelsep\usecounter{enumi}}}
 \relax
\def\reflist{\section*{References\markboth
        {REFLIST}{REFLIST}}\list
        {[\arabic{enumi}]\hfill}{\settowidth\labelwidth{[999]}
        \leftmargin\labelwidth
        \advance\leftmargin\labelsep\usecounter{enumi}}}
 \relax

\catcode`\@=11

\def\ps@headings{\def\@oddfoot{}\def\@evenfoot{}
\def\@oddhead{\hbox{}\hfill
        \makebox[.5\textwidth]{\raggedright\ignorespaces --\thepage{}--
        \hfill }}
\def\@evenhead{\@oddhead}
\def\subsectionmark##1{\markboth{##1}{}}
}

\ps@headings

\relax

\def\firstpage#1#2#3#4#5#6{
\begin{document}

\begin{titlepage}
\nopagebreak
\title{\begin{flushright}
        \vspace*{-1.8in}
        {\normalsize IC/95/177 -- NUB-#2\\[-9mm]CPTH--RR368.0795}\\[-9mm]
        {\normalsize hep-th/9507115}\\[4mm]
\end{flushright}
\vfill
{\large \bf #3}}
\author{\large #4 \\ #5}
\maketitle
\vskip -7mm
\nopagebreak
\begin{abstract}
{\noindent #6}
\end{abstract}
\vfill
\begin{flushleft}
\rule{16.1cm}{0.2mm}\\[-3mm]
$^{\star}${\small Research supported in part by\vspace{-4mm}
the National Science Foundation under grant
PHY--93--06906,\newline in part by the EEC contracts \vspace{-4mm}
SC1--CT92--0792 and CHRX-CT93-0340,
and in part by CNRS--NSF
grant INT--92--16146.}\\[-3mm]
$^{\dagger}${\small Laboratoire Propre du CNRS UPR A.0014.}\\
July 1995
\end{flushleft}
\thispagestyle{empty}
\end{titlepage}}
\newcommand{\dal}{\raisebox{0.085cm}
{\fbox{\rule{0cm}{0.07cm}\,}}}
\newcommand{\dt}{\partial_{\langle T\rangle}}
\newcommand{\dtbar}{\partial_{\langle\bar{T}\rangle}}
\newcommand{\al}{\alpha^{\prime}}
\newcommand{\mst}{M_{\scriptscriptstyle \!S}}
\newcommand{\mpl}{M_{\scriptscriptstyle \!P}}
\newcommand{\dv}{\int{\rm d}^4x\sqrt{g}}
\newcommand{\lv}{\left\langle}
\newcommand{\rv}{\right\rangle}
\newcommand{\ph}{\varphi}
\newcommand{\sbar}{\,\bar{\! S}}
\newcommand{\xbar}{\,\bar{\! X}}
\newcommand{\fbar}{\,\bar{\! F}}
\newcommand{\zbar}{\,\bar{\! Z}}
\newcommand{\tbar}{\bar{T}}
\newcommand{\ubar}{\bar{U}}
\newcommand{\ybar}{\bar{Y}}
\newcommand{\phb}{\bar{\varphi}}
\newcommand{\cm}{Commun.\ Math.\ Phys.~}
\newcommand{\pr}{Phys.\ Rev.\ D~}
\newcommand{\pl}{Phys.\ Lett.\ B~}
\newcommand{\ibar}{\bar{\imath}}
\newcommand{\jbar}{\bar{\jmath}}
\newcommand{\np}{Nucl.\ Phys.\ B~}
\newcommand{\e}{{\rm e}}
\newcommand{\gsi}{\,\raisebox{-0.13cm}{$\stackrel{\textstyle
>}{\textstyle\sim}$}\,}
\newcommand{\lsi}{\,\raisebox{-0.13cm}{$\stackrel{\textstyle
<}{\textstyle\sim}$}\,}
\date{}
\firstpage{95/XX}{3122}
{\large\sc N=2 Type II -- Heterotic Duality and\\[-4mm]
higher derivative F-terms$^{\star}$} {I. Antoniadis$^{\,a}$, E. Gava$^{b,c}$,
K.S. Narain$^{ c}$ $\,$and$\,$
T.R. Taylor$^{\,d}$}
{\normalsize\sl
$^a$Centre de Physique Th\'eorique, Ecole Polytechnique,$^\dagger$
F-91128 Palaiseau, France\\[-3mm]
\normalsize\sl
$^b$Instituto Nazionale di Fisica Nucleare, sez.\ di Trieste,
Italy\\[-3mm]
\normalsize\sl $^c$International Centre for Theoretical Physics,
I-34100 Trieste, Italy\\[-3mm]
\normalsize\sl $^d$Department of Physics, Northeastern
University, Boston, MA 02115, U.S.A.}
{We test the recently conjectured duality between $N{=}2$ supersymmetric
type II
and heterotic string models by analysing a class of higher dimensional
interactions in the respective low-energy Lagrangians.
These are $F$-terms of the form $F_g W^{2g}$
where $W$ is the gravitational superfield. On the type II side
these terms are generated at the $g$-loop level and in fact are
given by topological partition functions of the twisted
Calabi-Yau sigma model. We show that on the heterotic side these terms arise
at the one-loop level. We study in detail a rank 3 example and show that the
corresponding couplings $F_g$ satisfy
the same holomorphic anomaly equations as in the type II case.
Moreover we study the leading singularities of $F_g$'s on the
heterotic side, near
the enhanced symmetry point and show that they are universal poles of order
$2g{-}2$ with coefficients that are given by the Euler number of the moduli
space of genus-$g$ Riemann surfaces. This confirms a recent conjecture
that the physics near conifold singularity is governed by $c{=}1$ string
theory at the self-dual point.}
\section{Introduction}

In the last months considerable progress has been achieved in the
non-perturbative understanding of string theories with $N=2$ space-time
supersymmetry. In particular the idea of type II- heterotic string
duality has been extended to the $N=2$ context and some explicit examples
of dual pairs have been proposed by Kachru and Vafa \cite{kv}. This gives
rise to
an exact prepotential for the vector multiplets on the heterotic string
side, therefore extending the field theory results of Seiberg and Witten
\cite{sw} to the string case. An
important aspect of this duality is that the dilaton of one model is
mapped to an ordinary $t$-modulus associated with the compactification of
the second model. Moreover, the dilaton in type II  belongs to a hypermultiplet
while the dilaton in heterotic string belongs to a vector multiplet. Using
the fact that vector multiplets and neutral hypermultiplets do not couple
to each other, this duality provides a very powerful
method for extracting
non-perturbative physics of one model from the perturbative
computations in the dual model and vice versa.

Let us start by reviewing the main features of $N=2$ type II and
heterotic strings. The type II
string is compactified on a Calabi-Yau threefold\footnote{We shall be
considering here only (2,2) symmetric compactifications for type II.}
which is characterised by
the two Betti numbers $h_{11}$ and $h_{12}$. In the type IIA model, $h_{11}$
gives the number of vector multiplets. Together with the graviphoton the
rank of the gauge group is $r=h_{11}+1$. The number of hypermultiplets
is $h_{12}+1$ where the extra 1 is accounted for by the dilaton. At the
perturbative level the gauge group is abelian $U(1)^r$ and there are no
charged matter fields. Since the dilaton belongs to a hypermultiplet, the
tree level prepotential is exact at the full quantum level. Moreover this
tree level prepotential can be computed exactly, {\it i.e}.\ including the
world-sheet instanton corrections, by using the mirror symmetry
\cite{candm, cand, yau}. A generic feature of the prepotential is
that it
has logarithmic singularities near the conifold locus in the moduli space
of the Calabi-Yau threefold \cite{candc}. The appearance of this
logarithmic singularity at the tree level remained a puzzle for some time.
However recently Strominger \cite{andy} proposed a resolution of the
singularity
as due to the appearance of massless hypermultiplets, corresponding to
charged black holes, at the
conifold locus. The logarithmic singularity in the prepotential is then
understood as a one-loop effect involving this massless black hole in the
internal line. For this proposal to work it is crucial that the dilaton,
whose expectation value provides the loop expansion parameter,
does not couple to the vector multiplets at the two-derivative level.

The $N=2$ heterotic string is compactified on $T^2\times K_3$ with different
possible embeddings of the spin connection in the gauge group, giving
rise to different four-dimensional models. The moduli associated with this
compactification again split into two classes: the ones in vector
multiplets and those in hypermultiplets. Let us denote their numbers by
$n_V$ and $n_H$ respectively. Contrary to
the type II case, the dilaton $S$ in heterotic string belongs to a vector
multiplet. Thus the total rank of the gauge group including the
graviphoton becomes $(n_V+2)$. The $n_V$ moduli in the vector multiplets
belong to
the coset $O(2,n_V)/O(2)\times O(n_V)$ modulo discrete identifications that
define the duality group. At the classical level this duality group is
$O(2,n_V;Z)$. At the generic points in the moduli space the gauge group is
abelian $U(1)^{n_V+2}$ and there are no charged massless states. However
there are complex codimension 1 surfaces where one of the $U(1)$'s is
enhanced to $SU(2)$, due to the appearance of two extra charged massless
vector multiplets. As a result, the perturbative correction to the
prepotential, which due to the $N=2$ non-renormalization theorem occurs only at
the one-loop level, develops a
logarithmic singularity near these surfaces \cite{afgnt,dkll}.
As a result the classical duality group $O(2,n_V;Z)$ gets modified at the
perturbative level \cite{afgnt}.
At the full non-perturbative level, from the analysis of Seiberg and Witten
\cite{sw} in the rigid case, this enhanced symmetry locus is expected
to split into several branches where non-perturbative states
corresponding to dyonic hypermultiplets become massless. Thus in the full
moduli space
including the dilaton $S$, the singular locus should split into several
branches which collapse only in the limit $S \rightarrow \infty$.

The candidate dual pairs must of course have the same number of vector and
hypermultiplets. This means that $h_{11}=n_V+1$ and $h_{12}=n_H-1$. Moreover
the singularity structure for the prepotentials discussed above must also
be compatible with each other. Since the dilaton $S$ of heterotic is
mapped to an ordinary (1,1) modulus, say $t$, of type II, this implies that
first of all in the limit $t\rightarrow\infty$ the type II prepotential
must go to the perturbative prepotential of the heterotic theory
including the right singularity at the enhanced symmetry locus of the
latter. Moreover at finite values of $t$, the conifold singularity
structure must agree with what one expects from Seiberg-Witten analysis.
In other words different branches of conifold singularity must come
together as $t\rightarrow\infty$.

The examples of dual pairs proposed by Kachru and Vafa satisfy the above
qualitative requirements. Moreover for some examples involving rank 3 and
4, more quantitive checks have been made and it has been shown that the
prepotential in the type II
theory in the limit $S\rightarrow\infty$
agrees with the perturbative prepotential in the corresponding heterotic
model \cite{klt,klm,agnta}. These checks indicate that
at least at the two derivative level, the quantum effective actions of the
vector multiplets for these models are equivalent. However to show that
the two string theories are equivalent, one must go beyond the two
derivative terms. In particular as shown in
\cite{bcov,agntt}, there is a class of higher derivative
$F$-terms of the form $F_g W^{2g}$, where $W$ is the $N=2$ gravitational
superfield, which are generated at genus $g$ in the type II theory.
These terms again should not receive further quantum corrections even at the
non-perturbative level due to the fact that the type II dilaton is in the
hypermultiplet \cite{agntt, vafa}. The couplings $F_g$'s were shown to be
proportional genus $g$ topological
partition functions of the twisted Calabi-Yau sigma model, and satisfy
certain recursion relations expressing the holomorphic anomaly of genus
$g$ partition function to that of lower genera \cite{bcov}. If the
duality is true at the string level and not just at the level of low
energy effective action then such terms must also be present in the
heterotic string theory. For genus 1, this coupling corresponds to $R^2$
term in the effective action, $R$ being the Riemann tensor, and this has
already been investigated previously in the heterotic string \cite{agng}
and the anomaly equation satisfied by this coupling has been derived. For
rank 3 case it was shown in ref.\cite{klt} that $F_1$'s of the type II and
heterotic models agree in the weak coupling limit.

Our aim here is to analyse the whole sequence of $F_g$'s for
all $g$. It turns out that in heterotic string the terms
$F_g W^{2g}$ already appear at one-loop level. In this paper, we compute
these terms at the one-loop level for the rank 3 case, and derive the
holomorphic anomaly equations they satisfy. We show that these anomaly
equations are the same as the ones in type II theory in the large $S$ limit.
Moreover we analyse the leading singularity structure of $F_g$'s near the
enhanced symmetry point and show that the order of singularity matches
with that in type II. We also compute the coefficient of the leading
singularity of $F_g$ which gets contribution in the degeneration limit
of the world-sheet torus. Therefore this coefficient can be obtained by
calculating a one-loop diagram in effective field theory with the extra
massless states in the internal line.  It turns out that this coefficient is
the Euler number of the moduli
space of genus-$g$ Riemann surfaces, which is exactly the
coefficient of the $\mu^{2-2g}$ term in the expansion of the free
energy of $c=1$ string theory at the self-dual radius, $\mu$
being the cosmological constant. Although we do the explicit
calculation for the rank 3 case, it will be apparent that this
result is in fact universal. One would like to compare these
coefficients with the ones in the type II theory, where however one
has explicit results only up to $g=2$ (and for the quintic) which
agree with our results for heterotic string. On the other hand, Ghoshal
and Vafa \cite{gv} have recently argued that the physics near conifold
singularity is described precisely by the $c=1$ string theory at the
self-dual radius. If this turns out to be confirmed by further calculations
on the type II side, then our result would provide a strong evidence in favour
of the type II -- heterotic duality. Furthermore, the fact that these
singularities on heterotic side arise from a one-loop diagram with the
would-be massless state in the internal line, provides a strong
evidence in favour
of Strominger's proposal for the resolution of conifold singularity. In
other words in type II the leading singularity in $F_g$ should arise from
a one loop diagram involving the would-be massless charged black-hole in
the internal line.

This paper is organized as follows. In section 2 we discuss the $N=2$
effective field theory and show that while the couplings $F_g W^{2g}$ appear
in type II at $g$-loop level, in heterotic string they appear at one-loop
level. In section 3, we discuss the perturbative prepotential for
heterotic string for rank 3 case and compare with the type II. We also
discuss the type II holomorphic anomaly equations for $F_g$'s in the limit
$S\rightarrow\infty$. In section 4, we compute the $F_g$'s for heterotic
string. The generating function for the $F_g$'s can be expressed in a
compact way as an integral over the moduli space of the world-sheet torus.
In section 5, we derive the holomorphic anomaly equations for $F_g$'
for heterotic string and show that they are identical to the ones in type II
in large $S$ limit. We also compute the leading singularity near the
enhanced symmetry point and show that it gives the Euler number of the
moduli space of genus-$g$ Riemann surfaces. Section 6 is devoted to some
concluding remarks. Finally in the Appendix we derive a generating
function needed for the computations is section 4.

\section{Effective field theory}

The couplings in the effective field theory of type II strings which reproduce
the topological partition function $F_g$ were studied in ref.\cite{agntt}. It
was shown that they correspond to the chiral $N=2$ Langrangian terms
\begin{equation}
I_g = \tilde{F}_g(X)\, W^{2g} ,
\label{W2g}
\end{equation}
where $W$ is the (weight 1) Weyl superfield \footnote{For a general
discussion of $N=2$ supergravity see ref.\cite{dlv}.},
\begin{equation}
W_{\mu\nu}^{ij}=F_{\mu\nu}^{ij}-
R_{\mu\nu\lambda\rho}\,\theta^i\sigma_{\lambda\rho}\theta^j+\dots,
\label{chirW}
\end{equation}
which is anti-self-dual in its Lorentz indices and antisymmetric in the indices
$i,j$ labeling the two supersymmetries;
$W^2 \equiv \epsilon_{ij}\epsilon_{kl}W_{\mu\nu}^{ij}W_{\mu\nu}^{kl}$.
$R_{\mu\nu\lambda\rho}$ is the anti-self-dual Riemann tensor, while
$F_{\mu\nu}^{ij}$ is the (anti-self-dual) graviphoton field strength
defined by the supersymmetry transformation of the gravitinos:
$\delta\psi^i_{\mu}=-\frac{1}{4}\sigma^{\lambda\rho}F_{\lambda\rho}^{ij}
\sigma_{\mu}\bar{\epsilon}_j+\dots$. Finally $\tilde{F}_g(X)$ is an analytic
function of the $N=2$ chiral superfields $X^I$ (of Weyl weight 1):
\begin{equation}
X^I=X^I + \frac{1}{2}\widehat{F}^{I}_{\lambda\rho}
\epsilon_{ij} \theta^i\sigma_{\lambda\rho}\theta^j + \dots ,
\label{chirX}
\end{equation}
where $\widehat{F}^{I}_{\lambda\rho}$ are the anti-self-dual vector boson field
strengths. The scalar component of $X^0$ corresponds to a constrained field;
the unconstrained physical scalars of vector multiplets -- the moduli -- are
parametrized by $Z^A\equiv X^A/X^0$. By fixing the superconformal gauge, the
scalar component of $X^0$ can be expressed in terms of the moduli K\"ahler
potential $K(Z,\bar{Z})$ according to the normalization choice of the
coefficient of the Einstein kinetic term $R$. In the $\sigma$-model
representation, this coefficient is set to $1/g_s^2$, where $g_s$ is the
four-dimensional string coupling constant, implying that
\begin{equation}
X^0={1\over g_s}e^{K/2}
\label{X0}
\end{equation}
Since any Lagrangian term in conformal supergravity has Weyl weight 2, it
follows that $\tilde{F}_g(X)$ in eq.(\ref{W2g}) is a homogenous function of
$X^I$'s of degree $2\!-\!2g$. Its lowest component can then be written as
\begin{equation}
\tilde{F}_g(X)=(X^0)^{2-2g}\tilde{F}_g(Z)=(g_s^2)^{g-1} e^{(1-g)K}
\tilde{F}_g(Z)\, . \label{Fg}
\end{equation}
Note that $\tilde{F}_0$ coincides with the $N=2$ prepotential $F$ while
$\tilde{F}_1$ gives the gravitational four-derivative $R^2$-type couplings.
In ref\cite{agntt} it was shown that $\tilde{F}_g$ is of the form:
\be
\tilde{F}_g = \alpha \beta^{g-1} F_g
\label{prop}
\ee
where $F_g$ is the topological partition function of the twisted
Calabi-Yau sigma model and $\alpha$ and $\beta$ are some undetermined moduli
and $g$ independent constants.
Although $F_g$'s are expected to be analytic functions of the moduli, they
become non analytic (for $g\ge 1$) due to the holomorphic anomaly in the BRS
current of the topological theory \cite{bcov}. In the context of the effective
supergravity, the holomorphic anomaly is a consequence of propagation of
massless particles which lead to non-locality in the gravitational sector of
the
effective action \cite{agntt}. This is to be contrasted with the gauge sector
which is local at generic values of the moduli space where all non abelian
gauge
symmetries are broken to $U(1)$ factors and there are no massless charged
hypermultiplets. The moduli dependence of $F_g(Z,{\bar Z})$ is governed by the
following equation (for $g\ge 2$):
\begin{equation}
\bar{\partial}_{\bar{A}}F_g=\frac{1}{2}\bar{C}_{\bar{A}\bar{B}\bar{C}}e^{2K}
G^{B\bar{B}}G^{C\bar{C}}\left( D_BD_CF_{g-1}+
\sum_rD_BF_r\cdot D_CF_{g-r}\right),
\label{5eq}
\end{equation}
where $D$ are K\"ahler covariant derivatives, and the Yukawa couplings
$C_{ABC}\equiv F_{ABC}$, $F$ being the tree level prepotential. For $g=1$,
the equation which governs the moduli dependence of $F_1$ is:
\begin{eqnarray}
\bar{\partial}_{\bar{A}}\partial_A F_1&=&-{\chi\over 24}G_{A{\bar A}}+
{1\over 2}C_{ABC}{\bar C}_{\bar{A}\bar{B}\bar{C}}e^{2K}
G^{B\bar{B}}G^{C\bar{C}} \nonumber \\
&=& \frac{1}{2}\left( 3+h_{11}-\frac{\chi}{12}\right) G_{A\bar{A}} -
\frac{1}{2}R_{A\bar{A}} \label{anomeq1}
\end{eqnarray}
where $\chi$ is the Euler number of the Calabi-Yau manifold. In the
second step we have used the special geometry relation and $R_{A\bar{A}}$
is the Ricci tensor.

To count string loops, we use the fact that $N=2$ conformal supergravity
forbids a dependence of $F_g$'s on matter hypermultiplets, generalizing the
known result on the absence of mixing between vector multiplets and
hypermultiplets to the case of higher weight interactions. Next, we note
that in type II strings the K\"ahler potential $K$ is independent of the
dilaton
since the latter belongs to a hypermultiplet. Equation (\ref{Fg}) then implies
that the term $I_g$ in (\ref{W2g}) can appear only at genus $g$. Its highest
component contains a genus $g$ amplitude of two gravitons and $2g-2$
graviphotons which was studied in ref.\cite{agntt} in order to identify $F_g$
with the topological partition function of Bershadsky et al.\ \cite{bcov}.
These
arguments extend the non-renormalization theorem for the $N=2$ prepotential
$F_0$ to all $F_g$'s \cite{vafa}. Therefore, by analogy to the reasoning of
ref.\cite{andy}, we expect that in type II strings, $F_g$'s are determined at
genus $g$ and should not receive any further perturbative or non-perturbative
corrections.

On the other hand, heterotic -- type II string-string duality implies that
$F_g$'s should also appear in heterotic $N=2$ compactifications. In this case,
however, the dilaton belongs to a vector multiplet and $N=2$ supergravity does
not forbid a dependence of $F_g$'s on the dilaton. Fortunately, in the weak
coupling limit, Peccei-Quinn symmetry of the axion allows at most a linear
dilaton dependence for the prepotential $F_0$ and the $R^2$ coupling $F_1$, for
which a constant axion shift gives total divergences. Moreover, the K\"ahler
potential
$K$ contains a
$\ln g_s^2$ term. From eq.(\ref{X0}), one now has that
$X^0$ is of order 1 in the string coupling and thus, from eq.(\ref{Fg}), all
$F_g$'s should appear at the one loop, with the exception of $F_0$ and $F_1$
which have also tree level contributions. In conclusion, for $N=2$
compactifications which have dual realizations as type II and heterotic string
theories, the one loop corrections to the effective
gravitational couplings $F_g$ (\ref{W2g}) in the heterotic theory should
agree with the corresponding genus $g$ couplings in the dual type II theory
to the order $(S-\bar{S})^0$.

\section{Perturbative prepotential for the rank 3 case}

One of the simplest type II -- heterotic dual pairs proposed by Kachru and
Vafa is for rank 3 case. The type II model is compactified on the Calabi-Yau
threefold $X_{12}(1,1,2,2,6)$ with Betti numbers $h_{1,1}=2$ and
$h_{1,2}=128$. Thus the number of vector multiplets is $2$ and that of
hypermultiplets including the dilaton is $128+1$. The
classical prepotential as
a function of the special coordinates $t_1$ and $t_2$ of the moduli space of
vector multiplets has been studied in ref.\cite{cand}. They find the
following expression for the Yukawa couplings
\be
F_{\alpha\beta\gamma}= F^0_{\alpha\beta\gamma} + \sum_{0\leq j,k\in
\bf{Z}}\frac{c_ {\alpha\beta\gamma}n_{jk}q_1^j q_2^k}{1-q_1^j q_2^k},
\label{yukawa}
\ee
where $q_{\alpha}=\exp(2\pi i t_{\alpha})$ and $F^0_{\alpha\beta\gamma}$ are
the interserction numbers with $F^0_{111}=4$, $F^0_{112}=2$ and $F^0_{122}
=F^0_{222}=0$. The coefficients $c_{111}=j^3$, $c_{112}=j^2k$,
$c_{122}=jk^2$,
$c_{222}=k^3$ and $n_{jk}$ are the instanton numbers, the first few of
which have been explicitly given in ref.\cite{cand}. Note that $t_1$ and
$t_2$ are special coordinates so that $F_{\alpha\beta\gamma}$ are just the
appropriate derivatives with respect to $t_1$ and $t_2$ of the prepotential
$F$. Due to the fact that the dilaton in type II theory belongs to a
hypermultiplet, this
prepotential is exact and does
not receive any quantum correction even at the non-perturbative level.

The dual to the above type II model proposed in ref.\cite{kv} is an $N=2$
heterotic model with rank 3. The scalar components
of the vector multiplets are the dilaton $S$ and a modulus $T$ which belongs
to the coset $O(2,1)/O(2)$. We
use the normalization such that
$\langle S \rangle =\frac{\theta}{\pi}+i\frac{8\pi}{g_s^2}$. The classical
duality symmetry acting on $T$ is
$O(2,1;Z)\equiv Sl(2,Z)$. At generic point in the $T$-moduli space,
the gauge group is abelian, namely $U(1)^3$ including the vector partner
of dilaton and the graviphoton. However at $T=i$ (mod $Sl(2,Z)$) two extra
vector multiplets become massless, giving rise to an enhanced gauge group
$U(1)^2 \times SU(2)$.
The prepotential for this model is given
by
\be
F=\frac{1}{2}ST^2 + f(T) +...~~~,
\label{prep}
\ee
where $f(T)$ is the one loop correction to the classical prepotential
$\frac{1}{2}ST^2$, and due to $N=2$ non-renormalization theorems is actually
the complete perturbative correction. The dots refer to non-perturbative
contributions, which are suppressed exponentially as $e^{2\pi i S}$. Due to the
appearance of extra charged massless states at $T=i$, we expect a logarithmic
singularity in the one-loop contribution $f$ to the prepotential.
One can construct the K\"ahler potential starting from the prepotential $F$
up to
order $1/(S-\bar{S})$ and the  result is
\be
K= -\ln [\frac{i}{2}(S-\bar{S})(T-\bar{T})^2] + \frac{2i}{S-\bar{S}}
K^{(1)}(T,\bar{T}) \label{kahler}
\ee
where
\be
K^{(1)} = \frac{i}{T-\bar{T}}(\partial_T -\frac{2}{T-\bar{T}}) f + c.c
\label{k1}
\ee

The requirement that the $Sl(2,Z)$ transformations of $T$ should
be K\"ahler transformations implies that $f(T)$
transforms with weight $(-4)$ up to additive terms that are at most
quartic in $T$. Moreover $S$ also picks up additive terms that depend on
$f(T)$. Under the transformation $T\rightarrow \frac{aT+b}{cT+d}$
\begin{eqnarray}
f&\rightarrow& (cT+d)^{-4}(f + R) \nonumber \\
S&\rightarrow& S +2c\frac{f_T + R_T}{cT+d} -4c^2 \frac{f+R}{(cT+d)^2}
-\frac{1}{3}(R_{TT}-R_2)
\label{sl2zt}
\end{eqnarray}
where $R$ is a polynomial with real coefficients which is at most quartic in
$T$ and $R_2$ is the coefficient of $T^2$ in $R$. The term involving $R_2$
in the transformation of $S$ represents a constant axion shift and has been
introduced here for convenience. One can construct the one loop correction
to the metric
by taking derivatives of $K^{(1)}$ with the result:
\be
K^{(1)}_{T\bar{T}}= \frac{i}{(T-\bar{T})^2}(\partial_T-\frac{2}{T-\bar{T}})
(\partial_T-\frac{4}{T-\bar{T}}) f + c.c.
\label{g1tt}
\ee
{}From the fact that at $T=i$ extra charged massless states appear it
follows that the one-loop metric must have a singularity of the form
$\ln |T-i|$ for $T$ close to $i$. This in turn impies that $f$ must
behave as $(T-i)^2 \ln (T-i)$.

Under $Sl(2,Z)$ duality transformation
the one-loop metric must transform covariantly. Using eq.(\ref{g1tt}) one
can show the following identity:
\be
(\partial_T + \frac{4}{T-\bar{T}})(\partial_T + \frac{2}{T-\bar{T}})
\partial_T(T-\bar{T})^2 K^{(1)}_{T\bar{T}} =
i\partial_T^5 f \, .
\label{difff}
\ee
The left hand side of the above equation transforms with weight $(6,0)$
with respect to $(T,\bar{T})$. The right hand side is meromorphic in $T$
with at most a third order pole at $T=i$. Moreover as $T\rightarrow
\infty$ we expect the right hand side to vanish as otherwise it
would imply that $K^{(1)}_{T\bar{T}}$ would grow as $(T-\bar{T})$. The most
general expression compatible with these requirements is
\be
\partial_T^5 f =-\frac{1}{18 \pi i}
\left\{\frac{j_T}{j-j(i)}\right\}^3\left\{\frac{j(i)}{j}\right\}^2
\left\{5+13\frac{j}{j(i)}\right\}\, ,
\label{difff1}
\ee
where $j\equiv j(T)$ is the unique $Sl(2,Z)$ invariant meromorphic function
with
a first order pole at infinity with residue 1 and a third order zero at
$T=\exp(2\pi i/3)$. The constants appearing inside the bracket on
the right hand side are fixed by $SU(2)$ beta function which determines
the singularity in $K^{(1)}_{T\bar{T}}$ near $T=i$ and by the
requirement that the monodromy of $f$ as $T$ goes around $i$
be imbeddable in the symplectic group $Sp(6,Z)$ as dictated by $N=2$
supergravity.

To check the duality between type II and the heterotic models, we must
compare the prepotentials in the two theories in the weak coupling limit.
To do that we have to identify the moduli $(t_1,t_2)$ appearing in the
type II theory with $(S,T)$ appearing in the heterotic theory. The
identification
proposed in ref.\cite{kv} is $T=t_1$ and $S=2 t_2$.
One can indeed verify in the large $S$ limit ({\it i.e}.\ ignoring the
terms exponentially small in $t_2$) that the two expressions for the
prepotentials agree up to the first few terms in $q_1$ expansion that have
been checked so far \cite{klt}. Thus the two models seem to agree
at least up to the two derivative terms in the effective action in
the vector multiplet sector.

As stated in the introduction our aim here is to establish this equivalence
for all higher weight F-term couplings of the type
$F_g W^{2g}$ where $W$ is the $N=2$ gravitational Weyl supermultiplet. In the
type II theory $F_g$'s have already been related to genus $g$ topological
partition function of the twisted version of the Calabi-Yau sigma model. As
described in
the last section these $F_g$'s satisfy the recursion relations
(\ref{5eq}, \ref{anomeq1}). In order
to compare these couplings with the similar ones in heterotic string theory
we must again consider the large $S$ limit, as in the heterotic string
theory these quantities will be computable only in perturbation theory.
In the large $S$ limit it is easy to see that $\exp (2K)$ in
eqs.(\ref{5eq}, \ref{anomeq1}) becomes
$-4(S-\bar{S})^{-2} (T-\bar{T})^{-4}$. In the leading term in
$(S-\bar{S})$, the
only Yukawa coupling that is relevant in the anomaly equation is
$C_{\bar{T} \bar{T} \bar{S}}$, which is equal to 1. The inverse metrics that
enter the equation are in the leading orders
\begin{eqnarray}
G^{T \bar{T}}&=&-\frac{1}{2} (T-\bar{T})^2 ~~~~~~~~~~G^{S \bar{S}}
=-(S-\bar{S})^2
\nonumber \\
G^{S \bar{T}}&=&-i(T-\bar{T})^2 K^{(1)}_T
\label{ginverse}
\end{eqnarray}
Finally noting that, in the large $S$ limit, $F_g$ for $g\geq 2$ go to
constant in $S$ while $F_1$ goes to $-\pi i(S-\bar{S})$ plus zeroeth
order in $S$, we
find that to the leading order in $S$ the recursion relation (\ref{5eq})
becomes
\be
\partial_{\bar{T}} F_g = \frac{2\pi i}{(T-\bar{T})^2} (D_T F_{g-1}
+2\pi\delta_{g,2} K^{(1)}_T ) ~~~ g\geq 2
\label{leadrec}
\ee
For $g=1$, to the leading order in $S$, the anomaly equation (\ref{anomeq1})
becomes
\be
\partial_T \partial_{\bar{T}} F_1 =  \frac{-25}{(T-\bar{T})^2}
\label{leadanom}
\ee
where we have used the fact that for the present model the number of
hypermultiplets is 129.

Note that taking derivatives of $F_g$'s with respect to ${\bar S}$, one finds
contributions which do not vanish exponentially in the large $S$ limit but they
fall off as powers. This implies that in the heterotic theory $F_g$'s receive
in general higher loop corrections which deserve further study. In this work,
we
restrict ourselves to the leading weak coupling limit of $F_g$'s. However, when
making the comparison between type II and heterotic theories, there is a
related
subtlety which arises from the fact that the natural string basis
for the dilaton in heterotic theory corresponds to a linear multiplet
$L$, while in the dual type II model it is associated to a chiral
multiplet $S$. The relation between the two fields is \cite{der}
\be
\frac{1}{L} = {\rm Im}S - K^{(1)}
\label{lineardil}
\ee
Thus, changing variables from $T$ and $S$ to $T$ and $L$, one finds that the
partial derivatives with respect to $T$ in the two cases are related as
\be
\partial_T |_L = \partial_T |_S + K^{(1)}_T \partial_{{\rm Im}S}
\label{partial}
\ee
In the following when we derive holomorphic anomaly equation for $F_g$'s
in heterotic string, the partial derivative with respect to $T$ will be
for fixed $L$ and therefore to compare with type II equations which are
for fixed $S$ we need to use the above equation.

\section{Computation of $W^{2g}$ couplings in Heterotic String}

As we argued in section 2, on the basis of non-renormalization
properties of type II strings, if type II- heterotic
duality is correct, the expression for $F_g$, which is purely the
result
of a $g$-loop computation on the type II side, has to agree, in the weak
coupling limit $S\rightarrow \infty$,  with the result of the perturbative
computation on the heterotic side. In particular the zeroeth order term in
$(S-\bar{S})$ on the type II side should agree with the one loop results
on the heterotic side. As mentioned in section 2, the precise
identification of the coupling $\tilde{F}_g$ and the topological
partition function $F_g$ involved some undetermined constants $\alpha$
and $\beta$ in eq.(\ref{prop}), therefore in the
heterotic string
amplitude corresponding to $F_g$ which we shall compute below, we will not
be careful about such constants appearing in various steps.
At the end however we will normalize the amplitudes by demanding
that the coefficient appearing in the recursion relation on the heterotic
side be identical to the one appearing in eqs.(\ref{leadrec}) and
(\ref{leadanom}). In the following, therefore,
we shall also drop the distinction between $\tilde{F}_g$ and $F_g$.

Consider the amplitude involving two gravitons and $(2g-2)$
graviphotons. The relevant terms in the action are obtained by expanding
$F_g W^{2g}$ in terms of component fields with the result:
\begin{equation}
S_{eff} = gF_g (R^2)(F^2)^g + 2g(g-1) F_g (RF)^2(F^2)^{g-2}
\label{seff}
\end{equation}
where $R^2=R_{abcd} R^{abcd}$, $F^2 = F_{ab}F^{ab}$ and
$(RF)^2 = (R_{abcd}F^{cd})(R^{abef}F_{ef})$. As mentioned previously,
it is understood that $R_{abcd}$ and $F_{ab}$ represent the anti-self-dual
parts of the Riemann tensor and graviphoton field strengths respectively.

The vertices for the anti-self-dual parts of these fields are more easily
expressed by going to a complex basis for the four dimensional Euclidean
space time. Let us define
\begin{equation}
Z^1=(X^1-iX^2)/\sqrt{2} ~~~~~~~ Z^2 = (X^0-iX^3)/\sqrt{2},
\label{zs}
\end{equation}
and similarly for their left moving fermionic partners
\begin{equation}
\chi^1=(\psi^1-i\psi^2)/\sqrt{2} ~~~~~~~ \chi^2 = (\psi^0-i\psi^3)/\sqrt{2},
\label{psis}
\end{equation}
Then using the results in Appendix of ref.\cite{agntt} it is easy to see
that the following vertices describe the self-dual parts of Riemann tensor:
\begin{eqnarray}
V_h(p_1)&=& (\partial Z^2 - ip_1\chi^1\chi^2)\dbar Z^2 e^{ip_1 Z^1}
\nonumber \\
V_h(\bar{p}_2)&=& (\partial \bar{Z}^1 -
i\bar{p}_2\bar{\chi}^2\bar{\chi}^1)\dbar
\bar{Z}^1 e^{i\bar{p}_2 \bar{Z}^2}
\label{grav}
\end{eqnarray}
Here we have chosen convenient kinematics with $p_1\neq 0$,
$p_2=\bar{p}_1=\bar{p}_2=0$ for the first vertex and $\bar{p}_2\neq 0$,
$p_1=p_2=\bar{p}_1=0$ for the second (as usual we are treating $p$ and
$\bar{p}$ as independent parameters).

By applying $N=2$ space-time supersymmetry transformations twice one can
construct the vertices for the graviphotons in the same kinematic
configurations. In the zero ghost picture these are
\begin{eqnarray}
V_F(p_1)&=& (\partial X - ip_1\chi^1 \Psi)\dbar Z^2 e^{ip_1 Z^1}
\nonumber \\
V_F(\bar{p}_2)&=& (\partial X -
i\bar{p}_2\bar{\chi}^2 \Psi)\dbar
\bar{Z}^1 e^{i\bar{p}_2 \bar{Z}^2}
\label{photon}
\end{eqnarray}
where $X$ is the complex coordinate of the left-moving torus $T^2$ and
$\Psi$ is its fermionic partner with $U(1)$ charge $+1$.

Consider now an amplitude $A_g$ involving one graviton and $(g-1)$
graviphotons with $p_1\neq 0$,
$p_2=\bar{p}_1=\bar{p}_2=0$ and the remaining graviton and $(g-1)$
graviphotons with $\bar{p}_2\neq 0$, $p_1=p_2=\bar{p}_1=0$.
This amplitude gets contribution from both the terms in eq.(\ref{seff})
and it is easy to show that
\begin{eqnarray}
A_g &=& \langle V_h(p_1) V_h(\bar{p}_2) \prod_{i=1}^{g-1} V_F(p^{(i)}_1)
\prod_{j=1}^{g-1} V_F(\bar{p}^{(j)}_2) \rangle
\nonumber \\
&=& (p_1)^2(\bar{p}_2)^2 \prod_{i,j=1}^{g-1} p^{(i)}_1 \bar{p}^{(j)}_2
(g!)^2 F_g.
\label{anfn}
\end{eqnarray}

In general this amplitude receives contribution from all the spin
structures and one must sum over all the spin structures weighted
by a factor half associated to GSO projection. However one can show that
the sum over even spin structures gives the same contribution as the odd
one. Thus the full amplitude can be evaluated in the odd spin structure
without the factor of half.

In the odd spin structure one of the vertex operators must be inserted
in $(-1)$-ghost picture due to the presence of a Killing spinor on the
world sheet torus, and one must also insert a picture changing operator
to take care
of the world-sheet gravitino zero mode. It is convenient to take one
of the graviphoton vertices in the $(-1)$-ghost picture which comes with
a fermion $\Psi$. Recalling that in the odd-spin structure the space-time
fermions $\chi^i$ and $\bar{\chi}^i$, as well as the internal fermions
$\Psi$ and $\bar{\Psi}$ associated with the left-moving torus $T^2$
have one zero-mode each, one concludes that the only contribution comes
from the term $e^{\phi} \bar{\Psi} \partial X$ of the picture changing
operator. Moreover the space-time fermion zero modes are soaked by the
fermionic part of the graviton vertices. From the remaining $(2g-3)$
graviphoton vertices in the $(0)$-ghost picture only the terms involving
$\partial X$ survive. Together with the $\partial X$ appearing in the
picture changing operator they provide a total of $(2g-2)$ $\partial X$'s
which contribute only through their zero modes.

Finally we are left with the correlation functions of space-time
bosons. First thing to note is that $\dbar Z^2$'s and $\dbar
\bar{Z}^1$'s appearing in the vertex operators cannot contract with
each other. The same observation holds for the mutual contractions
between $e^{ip_1 Z^1}$'s and $e^{i\bar{p}_2 \bar{Z}^2}$'s.
Thus $\dbar Z^2$'s and $\dbar\bar{Z}^1$'s must contract with $e^{i\bar{p}_2
\bar{Z}^2}$'s and $e^{ip_1 Z^1}$'s, respectively, bringing down the appropriate
powers of momenta.  Moreover since
the correlator $\langle \dbar Z  \bar{Z} \rangle$ is total derivative,
in order to get non-vanishing result each $e^{i\bar{p}_2
\bar{Z}^2}$ must contract with some $\dbar Z^2$ and the same holds for
$e^{ip_1 Z^1}$'s. Thus the momentum structure of this amplitude matches
with the eq.(\ref{anfn}) and $F_g$ is given by the following expression:
\begin{eqnarray}
F_g &=&-\frac{(4\pi i)^{g-1}}{4\pi^2} \frac{1}{(g!)^2}\int
\frac{d^2\tau}{\tau_2^3}\frac{1}{\bar{\eta}^3}
\langle \prod_{i=1}^g \int d^2x_i Z^1\dbar Z^2(x_i) \prod_{j=1}^g
\int d^2 y_j \bar{Z}^2 \dbar \bar{Z}^1 (y_j) \rangle \nonumber \\
&~& \sum_{\epsilon=0,\frac{1}{2}} C_{\epsilon}(\bar{\tau})\sum_{m \in
\bf{Z}+\epsilon} \sum_{n_1,n_2
\in \bf{Z}} (\frac{iP_L}{T-\bar{T}})^{2g-2} q^{\frac{1}{2}|P_L|^2}
\bar{q}^{\frac{1}{2}P_R^2}
\label{fn}
\end{eqnarray}
where $\tau$ is the Teichmuller parameter of the world-sheet torus, $q$ is
$e^{2\pi i \tau}$ and  $1/\bar{\eta}^3$ accounts for
the partition function of the two space-time right
moving bosons in the light-cone and one free boson corresponding to the
right moving $U(1)$ current. Note that we have normalized the amplitudes
by putting in the factor $-(4\pi i)^{g-1}/4$, in order to match the
coefficients appearing in the recursion relation for $F_g$'s to that of
eqs.(\ref{leadrec}) and (\ref{leadanom}),
as it will be shown in the following. The correlators inside $\langle
...\rangle$ are normalized correlators.
$P_L$, $\bar{P}_L$ and $P_R$ are the left and right moving momenta
corresponding to the charges under $U(1)^3$. Explicitly they are given in
terms of $n_1$, $n_2$ and $m$ as:
\begin{eqnarray}
P_L &=& \frac{i\sqrt{2}}{T-\bar{T}} (n_1 + n_2 \bar{T}^2 + 2m \bar{T} )
\nonumber \\
P_R &=& \frac{i\sqrt{2}}{T-\bar{T}} (n_1 + n_2 T\bar{T} + m (T+\bar{T}) )
\label{mom}
\end{eqnarray}

Note that while $P_L$ is complex as they are the momenta associated with
the left-moving two-torus, $P_R$ is real. This is so because we are
considering the two moduli $(S,T)$ example and hence there is only
one $U(1)$ from the right-movers. The classical duality group that leaves
the spectrum invariant is $O(2,1)$ and the associated invariant inner product
is $\frac{1}{2}(P_L\bar{P}_L'+\bar{P}_L P_L') - P_R P_R' = (2mm'-n_1 n_2'
-n_2
n_1')$. The charges for the vector multiplets sit in a lattice $\Gamma_0$
defined by $n_1, n_2, m \in \bf{Z}$ and is even and integral. However
this lattice is not self dual with respect to the above inner product.
World-sheet modular invariance requires that the extra vectors contained in
the dual lattice $\Gamma_0^*$ given by $n_1, n_2 \in \bf{Z}$ and $m \in
\bf{Z}
+\frac{1}{2}$ must also appear in the full string spectrum. In fact
$\Gamma_0^*$ has two classes with respect to $\Gamma_0$ which we label
by $\Gamma_{\epsilon}$ for $\epsilon = 0,\frac{1}{2}$, with
$\Gamma_{\frac{1}{2}}$ being defined by $m \in \bf{Z} +\frac{1}{2}$. These
two classes couple to different blocks of the remaining conformal field
theory whose contribution to the above amplitude is denoted by $C_{\epsilon}
(\bar{\tau})$. Actually $C_{\epsilon}$ is the trace of $(-1)^F q^{L_0-c/24}
\bar{q}^{\bar{L}_0-\bar{c}/24}$ in the Ramond sectors of the corresponding
conformal blocks. They should only depend on $\bar{\tau}$. This can be argued
as follows. Since we are taking the trace of $(-1)^F$ in the Ramond sector
the non-zero modes of the left moving fermions must cancel exactly with
the left moving bosons. The only possible $\tau$ dependence can come from
instanton contributions. However for large K\"ahler class of the internal
$K_3$ surface these contributions if any would vanish. Changing the K\"ahler
class amounts to turning on vacuum expectation value for the corresponding
moduli which belong to hypermultiplets. Since the couplings $F_g$ do not
depend on hypermultiplets we conclude that these instanton contributions
must vanish also for finite K\"ahler class and hence $C_{\epsilon}$ must
depend only on $\bar{\tau}$. Note that the world sheet modular invariance
together with the existence of the tachyon in the right-moving sector
implies that
\begin{eqnarray}
C_0 &=& \bar{q}^{-7/8} \sum_{0\leq k \in \bf{Z}} a_k
\bar{q}^k ~~~~~~~~~~~ a_0=1,~~ a_1= -129 \nonumber \\
C_{\frac{1}{2}} &=& \bar{q}^{-\frac{5}{8}}
\sum_{0\leq k\in\bf{Z}} b_k \bar{q}^k
\label{fbehav}
\end{eqnarray}
Here $a_0=1$ accounts for the tachyon and $a_1 =-n_H$ where $n_H$ is the
number of hypermultiplets. In the case at hand $n_H=129$. This value of $a_1$
is
fixed by the requirement that $F_1$ reproduces the correct anomaly coefficient.

At the enhanced symmetry point $T=i$, one indeed finds two extra
massless states given by $n_1=n_2=\pm 1$ and $m=0$, which
enhance the gauge symmetry to $SU(2)\times U(1)^2$. In the
$\epsilon=0$ class there are no other points in the fundamental domain
of $T$ where there are extra massless states. However in the
$\epsilon=1/2$ class, at $T=\exp(2\pi i/3)$ one could get extra
charged massless states corresponding to $n_1=n_2=2m=\pm 1$ if
the coefficient $b_0$ in eq.(\ref{fbehav}) is not equal to zero.
However in this model one knows that there are no extra massless
states at $T=\exp(2\pi i/3)$ and therefore we conclude that
$b_0 = 0$. Furthermore from the knowledge of the modular transformation
properties of the lattice partition functions one can show that
$C_{\epsilon}$ transform under $\tau\rightarrow -1/\tau$ as
\be
C_{\epsilon} \rightarrow -\frac{1}{\sqrt{2}}(C_0 + e^{2\pi i \epsilon}
C_{\frac{1}{2}})
\label{fepsmod}
\ee

The correlation functions $\frac{1}{(g!)^2} \langle
\prod_{i=1}^g \int d^2x_i Z^1\dbar Z^2(x_i) \prod_{j=1}^g
\int d^2 y_j \bar{Z}^2 \dbar \bar{Z}^1 (y_j) \rangle$ appearing in
eq.(\ref{fn}) are just normalized free field correlators of space-time
bosons. In order to evaluate these correlation functions it is
convenient to introduce the following generating function:
\begin{eqnarray}
G(\lambda,\tau,\bar{\tau})&=&\sum_{g=1}^{\infty}\frac{1}{(g!)^2}
(\frac{\lambda}{\tau_2})^{2g} \langle \prod_{i=1}^g \int d^2x_i
Z^1\dbar Z^2(x_i) \prod_{j=1}^g
\int d^2 y_j \bar{Z}^2 \dbar \bar{Z}^1 (y_j) \rangle \nonumber \\
&\equiv & \sum_{g=1}^{\infty} \lambda^{2g} G_g (\tau,\bar{\tau})
\label{glamda}
\end{eqnarray}
The coeffecient $G_g$ times $\tau_2^{2g}$ is
what appears in the expression for $F_g$ (\ref{fn}). Note that
under the world-sheet modular transformation $\tau \rightarrow
\frac{a\tau + b}{c\tau+d}$ the $G_g$ transforms
with weight $2n$ in $\bar{\tau}$. Thus by assigning the transformation
$\lambda \rightarrow \frac{\lambda}{c\bar{\tau} +d}$, $G$ becomes invariant.
in other words
\be
G(\frac{\lambda}{c\bar{\tau} +d},\frac{a\tau + b}{c\tau+d},
\frac{a\bar{\tau} + b}{c\bar{\tau}+d}) =
G(\lambda,\tau,\bar{\tau})
\label{gmod}
\ee
Now $G$ can be expressed as the following normalized functional integral
of four bosonic fields:
\be
G(\lambda,\tau,\bar{\tau})=\frac{\int\prod_{i=1,2} DZ^i D\bar{Z}^i
exp (-S + \int \frac{\lambda}{ \tau_2} (Z^1\dbar Z^2 + \bar{Z}^2 \dbar
\bar{Z}^1 ) d^2 x ) }{\int\prod_{i=1,2} DZ^i D\bar{Z}^i
exp (-S)}
\label{gint}
\ee
The action $S$ is the free field action $S= \sum_{i=1,2}\frac{1}{\pi} \int
d^2 x (\partial Z^i \dbar \bar{Z}^i + \partial \bar{Z}^i \dbar Z^i)$.
Note that $\frac{1}{(g!)^2}$ appearing in eq.(\ref{glamda}) is
exactly taken care of in eq.(\ref{gint}). The right hand side of
this equation can be easily evaluated since the functional integrals
are gaussian. One can use $\zeta$-function regularization as in
ref.\cite{pol} to evaluate these functional integrals. In Appendix A we show
that the result of these functional integral gives the following
simple formula for $G$:
\be
G(\lambda,\tau,\bar{\tau})= (\frac{2\pi i\lambda \bar{\eta}^3}{\bar{\Theta}_1
(\lambda,\bar{\tau})})^2 e^{- \frac{\pi \lambda^2}{\tau_2}}
\label{gformula}
\ee
where $\Theta_1 (z,\tau)$ is the odd theta-function. This formula for $G$
can be understood as follows. The term involving
$\lambda$ in eq.(\ref{gint}) is just the right moving part of the space-time
Lorentz current. This term therefore effectively twists the boundary
conditions of the bosons by $e^{\pm2\pi i\lambda}$. This explains the
appearance of $\bar{\Theta}_1 (\lambda, \bar{\tau})^{-2}$ in
eq.(\ref{gformula}) since we have four bosons. $\bar{\eta}^6$ appears
because we are considering normalized correlators. $\exp{(-
\frac{\pi \lambda^2}{\tau_2})}$ is just due to the shift in the zero
point energy of the twisted bosons. In fact this can also be deduced
from the modular transformation of $\Theta$-function and the modular
invariance of $G$ following from eq.(\ref{gmod}). The appearance of
$\lambda^2$ also follows from modular invariance of $G$. The fact that
the right-moving Lorentz current is not a dimension (1,1) conformal
operator does not create any problem, since we are integrating over
flat world sheet torus. Indeed we have explicitely verified
eq.(\ref{gformula})
for $g=1,2,3,4$ by direct evaluation of the leading behaviour
in $1/\tau_2$ of the correlation functions (\ref{glamda}). These
leading terms also turn out to be crucial in studying the leading
singularities of $F_g$ near $T=i$, as we shall see below.

The generating function $G$ satisfies the following
differential equation which can be easily seen from eq.(\ref{gformula})
\be
\partial_{\tau} G(\lambda,\tau,\bar{\tau}) = -\frac{i\pi}{2}\frac{\lambda^2}
{\tau_2^2} G(\lambda,\tau,\bar{\tau})
\label{ganomaly}
\ee
This equation turns out to be important in evaluating the holomorphic
anomaly of $F_g$'s. In terms of the coefficients $G_g$ defined in
eq.(\ref{glamda}), and which appear in the definition of $F_g$
(\ref{fn}), this equation reads as:
\be
\partial_{\tau} G_g  = -\frac{i\pi}{2} \frac{1}{\tau_2^2} G_{g-1}
\label{bnanomaly}
\ee
Finally we close this section by giving an
expression
for the generating function of $F_g$'s in terms of world-sheet integral.
Define the following generating function
\be
F(\lambda, T,\bar{T}) = \sum_{g=1}^{\infty} \lambda^{2g} F_g
\label{flamda}
\ee
Then using the fact that $\lambda^{2g}$ appears with $\frac{1}{\tau_2}
(i\tau_2 P_L/(T-\bar{T}))^{2g-2}$ as follows from eqs.(\ref{fn})
and (\ref{glamda}), and using the explicit formula for $G$ (\ref{gformula}),
we can write the following expression for $F(\lambda,T,\bar{T})$:
\be
F(\lambda, T,\bar{T})= -\frac{1}{4\pi^2}\int
\frac{d^2\tau}{\tau_2}\frac{1}{\bar{\eta}^3}
\sum_{\epsilon=0,\frac{1}{2}} C_{\epsilon}(\bar{\tau})\sum_{m \in
\bf{Z}+\epsilon} \sum_{n_1,n_2 \in \bf{Z}}
(\frac{2\pi i\lambda \bar{\eta}^3}{\bar{\Theta}_1
(\tilde{\lambda},\bar{\tau})})^2 e^{-\frac{\pi}{2}
\tilde{\lambda}^2 \tau_2} q^{\frac{1}{2}|P_L|^2}
\bar{q}^{\frac{1}{2}P_R^2}
\label{flamint}
\ee
where $\tilde{\lambda}=\sqrt{-4\pi i} \lambda P_L/(T-\bar{T})$.

\section{Holomorphic anomaly and the leading singularity of $F_g$}

In this section we are going to use the results of section 4 to
perform the tests of heterotic-type II duality we have promised in the
introduction. First, we are going to compare the recursion relations
obeyed by the $F_g$'s, as computed in the previous section, to those of the
type II side in the $S\rightarrow \infty$ limit. Second, we will compute
the leading infrared singularity in the $F_g$'s near the enhanced
symmetric point $T=i$, and compare it with what one expects from the type
II side.

Recall that in
terms of the functions $G_g$, the couplings $F_g$ are expressed as:
\be
F_g = - \frac{(4\pi i)^{g-1}}{4\pi^2}\int
\frac{d^2\tau}{\tau_2^3}\frac{\tau_2^{2g}}{\bar{\eta}^3}
G_g (\tau,\bar{\tau}) \sum_{\epsilon=0,\frac{1}{2}}
C_{\epsilon}(\bar{\tau})\sum_{m \in \bf{Z}+\epsilon} \sum_{n_1,n_2
\in \bf{Z}} (\frac{iP_L}{T-\bar{T}})^{2g-2} q^{\frac{1}{2}|P_L|^2}
\bar{q}^{\frac{1}{2}P_R^2}
\label{fnbn}
\ee
We now wish to find the holomorphic anomaly equation satisfied
by $F_g$ and relate it to the corresponding equations in type II
theory. Let us take the derivative of $F_g$ with respect to $\bar{T}$.
One can prove the following identities which follow from the definitions
of $P_L$ and $P_R$ in eq.(\ref{mom}):
\begin{eqnarray}
&~&\partial_{\bar{T}}(\frac{P_L}{T-\bar{T}}) = 2 \frac{P_R}{(T-\bar{T})^2}
\nonumber \\
&~&\partial_{\bar{T}}(\bar{P}_L(T-\bar{T})) = 0
\label{identity}\\
&~&\partial_{\bar{T}}P_R =  \frac{\bar{P}_L}{T-\bar{T}}
\nonumber
\end{eqnarray}
There are of course similar identities for derivatives with respect
to $T$ which are just the complex conjugate of the above. Using these
identities one can easily show that for $g \geq 2$,
\begin{eqnarray}
\partial_{\bar{T}} F_g =\frac{i}{4\pi^3}\frac{(4\pi i)^{g-1}}{(T-\bar{T})^2}
\int d^2\tau  G_g (\tau,\bar{\tau})\,\times &&\label{fnam}\\ &&
\hspace*{-3.5cm}\partial_{\tau}
\bigl{[}\frac{\tau_2^{2g-3}}{\bar{\eta}^3}\sum_{\epsilon=0,\frac{1}{2}}
C_{\epsilon}(\bar{\tau})\sum_{m \in \bf{Z}+\epsilon} \sum_{n_1,n_2
\in \bf{Z}} (\frac{iP_L}{T-\bar{T}})^{2g-4} \partial_T
(q^{\frac{1}{2}|P_L|^2} \bar{q}^{\frac{1}{2}P_R^2})\bigr{]}
\nonumber
\end{eqnarray}
Now we can perform a partial integration with respect to $\tau$. The boundary
term vanishes for generic values of $T$ away from the singularity $T=i$. The
only nonvanishing contribution then appears when $\partial_{\tau}$ acts on
$G_g$. Using now eq.(\ref{bnanomaly}) one obtains
\be
\partial_{\bar{T}} F_g =  \frac{2\pi i}{(T-\bar{T})^2} D_T F_{g-1}
\label{fnanom1}
\ee
where $D_T$ is the K\"ahler covariant derivative. Recalling that
$F_g$ has K\"ahler weight $(g-1)$, which implies that it transforms as
weight
$(2g-2)$ with respect to $T$, one has the following action of $D_T$
\be
D_T F_g = (\partial_T - (g-1)(\partial_T K))F_g = (\partial_T +
\frac{2g-2}{T-\bar{T}}) F_g
\label{covder}
\ee
For $g=1$, the anomaly equation has been derived before in ref.\cite{agng},
or one can alternatively derive it using eqs.(\ref{fnbn}) and
(\ref{identity}), and the result is \begin{eqnarray}
\partial_{\bar{T}}\partial_T F_1 &=& \frac{25}{2} K^{(0)}_{T\bar{T}}
+\frac{2i}{\pi(T-\bar{T})^2}
\int \frac{d^2 \tau} {\tau_2^{3/2}} \frac{1}{\bar{\eta}^3}
\sum_{\epsilon=0,\frac{1}{2}}
C_{\epsilon}(\bar{\tau})\sum_{m \in \bf{Z}+\epsilon} \sum_{n_1,n_2
\in \bf{Z}} \partial_{\bar{\tau}}( \tau_2^{1/2}q^{\frac{1}{2}|P_L|^2}
\bar{q}^{\frac{1}{2}P_R^2}) \nonumber \\
&=& \frac{25}{2} K^{(0)}_{T\bar{T}} + 2\pi K^{(1)}_{T\bar{T}}
\label{f1anom}
\end{eqnarray}
where $K^{(0)}_{T\bar{T}}$ and $K^{(1)}_{T\bar{T}}$ are the tree level
metric and one loop correction to the metric respectively.
The first term on the right hand side comes from the boundary term
as $\tau_2 \rightarrow \infty$, and the second term appears as
a result of partial integration and the fact that $\partial_{\tau} G_1
= \frac{-i\pi}{2\tau_2^2}$. In the second step, we have used the fact that
the second term on the right hand
side is in fact just the Green-Schwarz term, which is proportional to the one
loop correction to the K\"ahler metric namely $K^{(1)}_{T\bar{T}}$

Now we wish to compare these results with the anomaly equations for
type II couplings $F_g$, in the leading $S$ limit eq.(\ref{leadanom}).
Note that
the coefficient of tree level metric $K^{(0)}_{T\bar{T}} = -2/(T-\bar{T})^2$
agrees in
the two equations (\ref{leadanom}) and (\ref{f1anom}). The appearance
of the extra term $K^{(1)}_{T\bar{T}}$ in (\ref{f1anom}) can be
understood as follows. As discussed in section 2, in the
type II case the anomaly equation is derived treating $T$ and $S$ as
independent variables, while in the heterotic string the independent
variables are $T$ and the linear dilaton multiplet $L$. Thus, in order to
compare the two equations, one must change the variables. As mentioned in
section 3, the linear
dilaton $L$ is related to $S$ via a duality transformation which gives
eq.(\ref{lineardil}). As a result, the partial derivatives with respect to
$T$ for
fixed $L$ and $S$ respectively are related by eq.(\ref{partial}). As noted
above only $F_1$, goes linearly as $\makebox{Im}(S)$ with
constant coefficient, while the remaining $F_g$'s for $g\geq 2$ have no
linear dependence on $S$. This means that the second
term on the right hand side of eq.(\ref{partial}) is non-trivial only
when it acts on $F_1$. It is then easy to see that
\be
\partial_T \partial_{\bar{T}} F_1 |_L = \partial_T \partial_{\bar{T}} F_1 |_S
{}~+~2\pi K^{(1)}_{T\bar{T}}
\label{f1ls}
\ee
Comparing now the equations (\ref{f1anom}), (\ref{leadanom}) and
(\ref{f1ls}), we find that the two anomaly equations do agree.

The anomaly equation for $F_g$'s (\ref{fnanom1}), is identical to the one for
the type II case (\ref{leadrec}) for $g\geq 3$. The reason is of course
that the second term on the right hand side of (\ref{partial}) vanishes
when acting on $F_{g}$ for $g\geq 2$. The only exception is for $g=2$, in
which case taking into account eq.(\ref{partial}), one finds that the
equation (\ref{fnanom1}) becomes
\be
\partial_{\bar{T}} F_2 |_S = \frac{2\pi i}{(T-\bar{T})^2}(\partial_T F_1 |_S
+ 2\pi  K^{(1)}_T)
\label{f2anom}
\ee
This equation again agrees with (\ref{leadrec}). Thus we conclude that
the anomaly equations for type II and heterotic strings agree at the
perturbative level.

We now turn to the question of holomorphic ambiguities in $F_g$'s. In
heterotic string, we have a closed form expression for $F_g$ as integral
over the moduli of the world-sheet torus. For type II on the other
hand the $F_g$'s involve integration of the topological partition
functions
of the twisted Calabi-Yau sigma models over the moduli space of genus-$g$
Riemann surfaces, and therefore the determination of holomorphic
ambiguities in this case is extremely difficult. However, one can try to
compare the leading singularities in $F_g$
near $T=i$. In the type II case, as noted in ref.\cite{bcov}, the leading
singularity in $F_g$ for $g\geq2$
near the conifold locus is $\mu^{2-2g}$, while for $F_1$ it is $\ln |\mu|$,
where $\mu$ is the local coordinate which goes to zero at the conifold.
These leading singularities are meromorphic and therefore are not
captured in the holomorphic anomaly equations. As $S\rightarrow \infty$,
the two branches of the conifold meet at $T=i$. Thus in this limit
we can identify $\mu$ with $(T-i)$, up to a constant multiplicative factor.
The coefficient of
this leading singularity is expected to be universal. This follows from
Strominger's interpretation
of the conifold singularity as due to the appearance of massless
charged black holes \cite{andy}. The singularity in $F_g$ then
would be due to a one-loop diagram involving this massless black
hole as the internal line. The universality follows from the fact that
the graviton and graviphoton couple universally to massless hypermultiplets.
It has also been argued by Ghoshal and Vafa \cite{gv}, that the leading
singularity structure of $F_g$'s is described by the free energy
of the $c=1$ string theory at the self dual radius:
\be
Z_{c=1}~=~ \frac{1}{2}\mu^2 \ln{\mu} -\frac{1}{12}\ln{\mu} + \sum_{g\geq
2}\chi(g) \mu^{2-2g}
\label{ceq1}
\ee
where $\chi(g)$ is the Euler number of the moduli space of genus-$g$
Riemann surfaces and $\mu$ is the cosmological constant. The identification
with the singularities of $F_g$ follows from identifying $\mu$ with a
local coordinate near conifold which vanishes at the conifold. The
normalization of the coordinate is fixed by comparing the tree level
singularity with that in eq.(\ref{ceq1}). In ref\cite{gv}, it was shown
that with this normalization, the coefficient of the singularity for genus
$g=2~$ is exactly $\chi(2)$ for the type II string compactified on the
quintic threefold.

In heterotic case also we expect the same singularity structure near $T=i$.
Here however, two elementary string states corresponding to charged vector
multiplets become massless and as a result a string one-loop computation
should exhibit this singularity structure. The universality of the
coefficient
of the leading singularity follows from the same argument as in the type II
case. We now evaluate these coefficients from the explicit expression
eq.(\ref{fnbn}) for $F_g$'s in the heterotic string. Near $T=i$, the extra
massless states correspond to the lattice states with $n_1=n_2=\pm 1$ and
$m=0$ in eq.(\ref{mom}). In this limit the left and the right moving
momenta $P_L$ and $P_R$ behave as
$P_L =-i\sqrt{2} (\bar{T}+i)$ and $P_R^2 =2 + 2|T-i|^2$. The singularity
appears
from the region of integration corresponding to large $\tau_2$. To perform
this integration and isolate the singularity it is convenient to rescale
$\tau_2$ as $4\pi\tau_2 |T-i|^2$. Taking into account the powers of $\tau_2$
and $P_L$ appearing in eq.(\ref{fnbn}), one finds that this change of
variable brings about a factor of $(4\pi\sqrt{2}(T-i))^{2-2g}$ which is
exactly
the expected leading singular behaviour. Furthermore only the constant term
in $~(\bar{q} \bar{\eta}^{-3}C_{0}G_g)~$ as $\tau_2 \rightarrow \infty$
contribute to the coefficient of singularity.
In this limit, $\bar{q} \bar{\eta}^{-3}C_{0} = a_0=1$, as follows from
eq.(\ref{fbehav}). Finally the leading term in $G_g$ is simply
the coefficient of $\lambda^{2g}$ in the expansion of
\be
\int d\tau_2 \tau_2^{2g-3} (\frac{\pi \lambda}{\sin \pi\lambda})^2
e^{-\tau_2} \label{schwinger}
\ee
We can compute this as follows:
\begin{eqnarray}
(\frac{\pi \lambda}{\sin \pi\lambda})^2 &=& -2\pi i\lambda^2
\partial_{\lambda} \frac{1}{e^{2\pi i\lambda}-1} \nonumber \\
&=&-\frac{(2g-1)(-1)^g}{(2g)!}B_{2g} (2\pi \lambda)^{2g}
\label{euler}
\end{eqnarray}
where $B_{2g}$ is the $2g$-th Bernoulli
number, and in the second step we have used the definition of the generating
function for Bernoulli numbers. For $g=1$, the integral over $\tau_2$ has
a logarithmic divergence near $T=i$, and taking into account the
value of the Bernoulli number $B_2 = 1/6$, one finds that $F_1$
behaves as $\frac{1}{6} \ln(T-i)$. Note that the coefficient of the
logarithmic singularity is
$(-2)$ times the one appearing near the conifold singularity in
eq.(\ref{ceq1}), namely
$(-1/12)$. Finally for $g\geq 2$, the integral $\int d\tau_2 \tau_2^{2g-3}
\exp(-\tau_2)$ provides an extra factor of $(2g-3)!$. Thus altogether
for $g\geq 2$,
the coefficient of the leading singularity
$(\sqrt{\frac{2}{\pi i}}(T-i))^{2-2g}$ is
$-2 B_{2g}/2g(2g-2)$ which is just $(-2)$ times
the Euler number $\chi(g)$  of the moduli space of genus-$g$ Riemann
surfaces. Thus by identifying $\mu$ with $\sqrt{\frac{2}{\pi i}}(T-i)$,
we find that the singularity structure for $F_g$'s near $T=i$ is
described by $(-2)$ times the free energy of $c=1$ string theory
at the self-dual point. In fact, with this identification, the tree level
term $-2(\frac{1}{2}\mu^2\ln{\mu})=\frac{2i}{\pi}(T-i)^2\ln(T-i)$, exactly
reproduces the singularity of
the prepotential $f$ discussed in section 3. The relative factor of $(-2)$
can be understood
from the fact, that while near $T=i$ one has two extra massless vector
multiplets, near the conifold only one hypermultiplet corresponding to a
black hole becomes massless. The ratio of their contributions to the trace
anomaly in the two cases respectively is exactly $(-2)$. Moreover this also
supports the argument of
ref.\cite{gv} that the physics near conifold singularity is described
by $c=1$ string theory at the self-dual radius. A more invariant
identification
of $\mu$, which applies generically to the singularities associated with the
appearance of massless states, is given by $\mu = \sqrt{i/\pi}
\exp(-K^{(0)}/2) Z$, where $Z$ is the central charge of the $N=2$
supersymmetry algebra \cite{cafv} and $K^{(0)}$ is the tree level K\"ahler
potential corresponding to the $T$ moduli. Note that $\mu$ transforms
with weight 1 under K\"ahler transformations, however recalling that
$\mu^{2-2g}$ appears together with $(2g-2)$ graviphotons, the
corresponding term in the effective action is K\"ahler invariant, due to the
transformation properties of the graviphotons discussed in section 2.

It is interesting to
note that a one loop computation in heterotic string reproduces the Euler
number of moduli space of genus-$g$ Riemann surfaces. In fact since the
leading singularity appears from $\tau_2 \rightarrow \infty$ limit, it
should be possible to understand it purely at the effective field theory
level and this in the type II context would then be consistent with
Strominger's interpretation of the conifold singularity. In fact the
effective action for QED in the presence of constant electric and
magnetic fields has been computed long ago by Schwinger \cite{sch}. If
one considers self-dual background, then Schwinger's formula exactly
coincides with eq.(\ref{schwinger}) with the identification
$\lambda^2=F_{\mu\nu}^2$. It is interesting to note that QED
already computes the Euler number of the moduli space of genus-$g$
Riemann surfaces! However, we are concerned here with $N=2$
supergravity sector and Schwinger's formalism needs to be extended to
this case.

It should be pointed out that from the heterotic side,
using the general formula (\ref{fnbn}), one can also compute quite
easily the subleading singularities near $T=i$. This
involves expanding $P_L$ near $T=i$ as well as the generating function $G_g$
whose relevant part is obtained by expanding
$(\frac{\pi\lambda}{\sin\pi\lambda})
^2 \exp(-\frac{\pi\lambda^2}{\tau_2})$. It would be interesting
to compare also these subleading singularities with the ones in type II,
where however we do not have similar results at present. On the other
hand assuming duality, the fact that we
can compute the coefficients of all the poles in $F_g$, can help at least
partially in
fixing the holomorphic ambiguity on the type II side.

\section{Concluding remarks}

In this paper we examined the proposed $N=2$ type II-heterotic duality in a
class of higher derivative F-terms of the form $F_g W^{2g}$, $W$ being
the gravitational multiplet. While in type II side, they appear at $g$-loop
level and are exact at the quantum level, on the heterotic side, to the
leading order in the string coupling, they appear at the one-loop level.
We analysed in detail the rank 3 example and showed that to this order,
the holomorphic anomaly
equations for $F_g$'s  are identical for the two models. We also computed
the leading singularity near the enhanced symmetry point in the heterotic
string and showed that the corresponding coefficient is universal and is
given by the Euler
number of the moduli space of genus-$g$ Riemann surfaces.
Therefore if the conjecture of Ghoshal and Vafa, relating the
conifold singularity to the $c=1$ string theory at self-dual
radius, would receive more evidence for $g\geq3$, then our result
would represent a very strong argument in favour of the type
II-heterotic duality. Although we
have focussed here on the rank 3 example, one can easily extend the above
analysis to the rank 4 case.

There are several questions which need further investigation.
One of the issues is regarding corrections to $F_g$'s that are higher
order in $1/(S-\bar{S})$. This involves going beyond one loop on the
heterotic side. On the type II side, the open issue is the structure of
leading and subleading singularities near the conifold. Another issue
which has not been investigated so far is the comparison of the
hypermultiplet sectors of the two theories.

In conclusion, our analysis of higher-dimensional
effective Lagrangian interactions provides a very strong quantitative
evidence supporting the duality conjecture for
certain pairs of type II and heterotic superstring
models in four dimensions.
The most intriguing aspect of duality which emerges very clearly
from this work is the apparent
equivalence of physical effects which occur at different loop orders, or even
non-perturbatively, as viewed from dual descriptions. This goes
very far beyond our experience with low-energy quantum field theory,
and our intuition what is classical and what is quantum.
Uncovering the origin of
duality may indeed provide a clue to understanding the physical content
of superstring theory.\\[1cm]

{\bf Acknowledgements}
We would like to thank Ashoke Sen and George Thompson for many useful
discussions. I.A. acknowledges the hospitality of I.C.T.P. and T.R.T.\ ,
E.G. and K.S.N. acknowledge the hospitality of the Centre de Physique
Th\'eo\-rique at Ecole Polytechnique.
\vspace{2.cm}

{\large\bf Appendix}\setcounter{equation}{0}
\def\theequation{A.\arabic{equation}}

This appendix is devoted to the derivation of eq.(\ref{gformula}). We will
adopt
the $\zeta$-function regularization used in ref.\cite{pol} to evaluate
the determinant of a scalar field on a world sheet torus. The functional
integral in eq.(\ref{gint}) is quadratic in the scalar fields, so we can
evaluate it by expanding the $Z$'s into an orthonormal basis of
eigenfunctions of the scalar laplacian. Let us choose torus coordinates
$\sigma^1$, $\sigma^2$, with $0\leq \sigma^1, \sigma^2 \leq 1$, and the
corresponding metric to be given by $ds^2 = |d\sigma^1 + \tau d\sigma^2|^2$.
The orthonormal basis is given by $\phi_{n,m} = \frac{1}{\sqrt{\tau_2}}
\exp(2\pi i (n\sigma^1 + m\sigma^2))$ where $n,m \in \bf{Z}$. It is then
easy to see that as a result of the functional integration in the
numerator of eq.(\ref{gint}) we will get the following determinant:
\be
{\det}' \Delta = \prod_{(n,m)\ne (0,0)} (\frac{2\pi}{\tau_2^2})^2
[|n-m\tau|^4 -\lambda^2(n-m\tau)^2]
\label{det}
\ee
which defines the operator $\Delta$ in terms of its eigenvalues. To evaluate
(\ref{det}) it is useful to split $\Delta$ as $\Delta=\Delta^+ \Delta^-$,
where $\Delta^{\pm}$ have eigenvalues
\be
\lambda_{n,m}^{\pm} = \frac{2\pi}{\tau_2^2}^2
[|n-m\tau|^2 \pm\lambda(n-m\tau)]
\label{detpm}
\ee
We can then evaluate $\ln {\det}'\Delta^{\pm}$ following ref.\cite{pol}
by using $\zeta$-function regularization and converting the sum over $n$ into
an integral using the Sommerfeld-Watson transformation. The result is:
\begin{eqnarray}
\ln {\det}' \Delta^{\pm} &=& \lim_{s,\mu^2\rightarrow 0} \nonumber \\
&-& 2
\frac{d}{ds}\bigl{[} (\frac{2\pi}{\tau_2^2})^{-s}\int_c dz \sum_m \frac{e^{i\pi
z}}{2i\sin\pi z}[(z-m\tau_1)^2 +
m\tau_2^2 \pm \lambda (z-m\tau) + \mu^2]^{-s} + {\rm h.c.}\bigr{]}
\nonumber \\
&+& \frac{d}{ds}\bigl{[}(\frac{2\pi}{\tau_2^2})^{-s}\int_c
dz \sum_m [(z-m\tau_1)^2 +
m\tau_2^2 \pm \lambda (z-m\tau) + \mu^2]^{-s} + h.c.\bigr{]}
\nonumber \\
&-&(\frac{2\pi\mu^2}{\tau_2^2})^{-s}\ln(\frac{2\pi \mu^2}{\tau_2^2}).
\label{logdet}
\end{eqnarray}
The contour passes above the real axis from $+\infty +i\epsilon$
to $-\infty +i\epsilon$ and we have introduced a mass $\mu$ as an
infrared regulator. Notice that the sum over $m$ includes $m=0$. Let us
first do the computation for
$\Delta^+$. The first term in the bracket converges at $s=0$ and gives
\be
2\sum_{m=1}^{\infty}\ln(1-q^m) +\sum_{m=0}^{\infty}\ln(1-\bar{q}^m
e^{-2\pi i \lambda} )+\sum_{m=1}^{\infty}\ln(1-\bar{q}^m e^{2\pi i
\lambda}) + \ln(2\pi \mu^2) -\ln(\lambda)
\label{etapart}
\ee
Note that $\ln(2\pi\mu^2)$ term cancels between eq.(\ref{etapart}) and
the last term in eq.(\ref{logdet}). The second integral can be expanded in
powers of $\lambda$. After performing the sum over $m\ne 0$,
it turns out that only up to quadratic terms in $\lambda$ survive in the
limit $s\rightarrow 0$ with the result:
\be
-\frac{\pi}{3}\tau_2 + \frac{\pi}{2}\frac{\lambda^2}{\tau_2} +
i\pi\lambda.
\label{zeropoint}
\ee
For $m=0$ the result vanishes in the limit $s,\mu \rightarrow 0$.
Combining now the contribution $\Delta^-$ and taking into account
the normalization i.e. the partition function of four scalars, we
get the desired result eq.(\ref{gformula}).

\end{document}